\documentclass[fleqn,usenatbib]{mnras}

\usepackage[T1]{fontenc}
\usepackage{ae,aecompl}

\usepackage{graphicx}	
\usepackage{amsmath}	
\usepackage{amssymb}	
\usepackage{bm}               
\title{Cosmic Ray Driven Outflows in an
  Ultraluminous Galaxy}

\author[Fujita]{
Akimi Fujita,$^{1}$\thanks{E-mail: fujitaa@shinshu-u.ac.jp}
and Mordecai-Mark Mac Low$^{2}$\thanks{Email: mordecai@amnh.org}
\\
$^{1}$Faculty of Engineering, Shinshu University, Nagano, Nagano, Japan\\
$^{2}$Department of Astrophysics, American Museum of Natural History,
New York, New York, U.S.A.\\
}

\date{Accepted 2018 Marh 17}

\pubyear{2018}

\begin{document}
\label{firstpage}
\pagerange{\pageref{firstpage}--\pageref{lastpage}}
\maketitle

\begin{abstract}
In models of galaxy formation, feedback driven both by supernova (SN)
and active galactic nucleus (AGN) is not efficient enough to quench
star formation in massive galaxies. 
Models of smaller galaxies 
 have suggested that
cosmic rays (CRs) play a major role in expelling material from the
star forming regions by diffusing SN energy to the lower density
outskirts. We therefore run gas dynamical simulations of galactic
outflows from a galaxy contained in a halo with
$5\times10^{12}$~M$_{\odot}$ that resembles a local ultraluminous galaxy, 
including both SN thermal energy and a treatment of CRs using the same
diffusion approximation as \citet{SalemBryan2014}. 
We find that CR pressure drives a low-density bubble beyond
the edge of the shell swept up by thermal pressure, but the main
bubble driven by SN thermal pressure overtakes it later, which creates
a large-scale biconical outflow. 
CRs diffusing into the disk are unable to entrain its gas in the
outflows, yielding a mass-loading rate of only $\sim0.1~\%$ with
varied CR diffusion coefficients
We find no significant difference in mass-loading rates in SN driven outflows with or without CR pressure. Our simulations strongly suggest that it is hard to drive a heavily mass-loaded outflow with CRs from a massive halo potential, although more distributed star formation could lead to a different result.
\end{abstract}




\section{Introduction}
In the theory of galaxy formation, feedback in the form of galactic
outflows plays an important role in regulating star formation (SF), and thus
the structure and evolution of galaxies. Outflows not only heat and
ionize gas but can also drive gas and metal from SF regions to galactic haloes
and beyond. Recent cosmological hydrodynamic simulations have
highlighted the importance of suppressing SF with stellar feedback in
galaxies towards the low mass end of the distribution 
and with active galactic nucleus (AGN) feedback in galaxies towards
the high mass end to reproduce the observed, global properties of
galaxies \citep[e.g.][]{Vogelsberger2013,Marinacci2014}. 
At the same time, these simulations showed discrepancies that require further
investigation, one of which is overproduction of stars at the present
epoch in galaxies with halo masses
$M_{h}\gtrsim 10^{12}$~M$_{\odot}$, even with a very energetic form of AGN
feedback compared to previous studies \citep{Vogelsberger2014}. 
This mass, $\sim 10^{12}$~M$_{\odot}$, seems to be the transition mass at which the main process
 regulating SF is stellar or supernova (SN) feedback below and AGN
 feedback above. 
Cosmic rays (CRs) are suggested to be one of the
major players in feedback. 
However, models of SN feedback in massive galaxies larger than the Milky Way
 have only included thermal pressure to date, not cosmic ray 
 pressure \citep[e.g.][]{Fujita2009, Melioli2015}.

CRs can
accelerate gas away from the dense ISM 
without radiatively losing energy, unlike thermal
superbubbles\citep[]{Breitschwerdt91}, and they lose less energy to
adiabatic expansion than thermal gas.  
Recent three-dimensional (3D) hydrodynamic (HD) and
magnetohydrodynamic (MHD)
simulations have demonstrated that CR driven winds can smoothly
accelerate colder, diffuse gas out of galactic disks and efficiently regulate SF in dwarf and Milky Way-like
galaxies.
(\citealt{Uhlige012}, \citealt{Hanasz2013}, \citealt{Booth2013},
 \citealt{SalemBryan2014,Simpson2016}, \citealt{Girichidis2016}, \citealt{Martizzi2016},
\citealt{Pakmor2016}, \citealt{Ruszkowski2016}).

These works approximate CR transport 
with a variety of assumptions: 1)
streaming at the local sound speed \citep[HD:][]{Uhlige012,
  Wiener2016}; 2) self-confinement \citep[MHD:][]{Ruszkowski2016};
3) isotropic diffusion \linebreak 
\citep[HD:][]{Booth2013}, \citealt{SalemBryan2014}; or 4) anisotropic diffusion
with fixed diffusion coefficients \citep[MHD:][]{Hanasz2013,
  Girichidis2016}, \citealt{Pakmor2016,Ruszkowski2016}. 
\citet{Wiener2016} in their study comparing streaming to diffusion
point out that the dynamical effects of diffusion are overestimated by
an order of magnitude when wave excitation and the transfer of energy
from CRs 
to thermal gas by streaming are ignored in dwarf-sized galaxies, while
\citet[]{Ruszkowski2016} demonstrate that the dynamical effects of
streaming and diffusion are comparable 
if wave growth due to the streaming instability is inhibited by some
damping process, such as turbulent damping, even in a Milky Way-sized galaxy.  

The effects of CR pressure have not been studied in 
  galaxies more massive 
than the Milky Way. 
We therefore run simulations of galactic outflows, including both SN
 thermal energy and cosmic rays, 
  with varying isotropic coefficient of diffusion,
 in a massive galaxy that resembles a
 local ultraluminous galaxy  
    some $\sim5$ times
than the Milky Way. Our galaxy and galactic outflow models are 
based on \citet{Fujita2009} with a halo mass
 $5\times10^{12}$~M$_{\odot}$ and a mechanical luminosity $L_{mech}
 =10^{43}$\,erg\,s$^{-1}$
   provided by supernova explosions and associated stellar outflows,
   and treated as a continuous energy input \citep{ML1988}. 
Our previous study showed that 
SN thermal feedback alone is not effective in 
    driving a heavily mass-loaded outflow
from its deep potential. 
    Therefore, 
we compute the mass-loading rates of CR-driven outflows in the same
galaxy with the same SF and superbubble recipe by including a diffusive CR
 fluid, as described in \citet{SalemBryan2014}. 
We assume that AGN feedback still remains quiescent in the galaxy. 

In this paper, we describe our numerical methods as well as our disk
and outflow models in Section 2 and the mass-loading rates, as well as resolution studies in Section 3, followed by discussion in Section 4
and conclusion in Section 5. 

\section{Numerics}

\subsection{Algorithm}
The simulations were performed using the adaptive mesh refinement hydrodynamics code ENZO
\citep[e.g.][]{Bryanetal2014} with a two-fluid model for CRs \citep[]{DruryFalle86}
implemented and tested by \citet{SalemBryan2014}. In this model, CRs
are treated as a 
relativistic gas with $\gamma_{CR}=4/3$ that is coupled to the thermal
plasma with $\gamma_{th}=5/3$ except for a diffusion term. Bulk motions
of the thermal gas transport the CRs, and they in turn diffuse and
exert pressure on the thermal gas.  

This model assumes that CRs diffuse isotropically by
scattering off inhomogeneities in the tangled magnetic
fields near the disk midplane, with a fixed CR diffusion coefficient
$\kappa_{CR}$. Recent MHD simulations show that 
outflows develop later with anisotropic
diffusion, but only in cases with magnetic fields dominated by azimuthal
components \citep[]{Pakmor2016}, and that anisotropic diffusivities
act similar to isotropic diffusivities with $\kappa_{CR}\sim
\kappa_{||}/3$ where $\kappa_{||}$ is a fixed CR diffusion coefficient
parallel to the magnetic fields \citep[]{Ruszkowski2016}. 
In both studies, the global mass-loading rates with anisotropic diffusion are
comparable to those with isotropic diffusion. 

We apply a ceiling to the effective sound speed of
$c_{s,max}=2.05 \times 10^4$~km~s$^{-1}$ in low-density, CR-dominated
cavities by raising the density within them 
in the highest resolution run (QC: see Table~\ref{tab:table1}). 
This is necessary to
prevent our computation from slowing down significantly, but only a
very small portion of the gas feels this effect, causing no
significant change in the dynamics \citep{SalemBryan2014}. 

\subsection{Cosmic Ray Parameters}
We choose values of the diffusion coefficient
$\kappa_{CR}=3\times10^{27}$\,cm$^{2}$\,s$^{-1}$, which yielded the highest
mass-loading rate of $\sim1.0$ in \citet{SalemBryan2014}, and an order
of magnitude higher and lower. \citet{SalemBryan2014} found that low
diffusion rates prevent CRs from diffusing out of the densest regions
quickly enough to 
create the pressure gradient in the upper atmosphere of the disk
required to accelerate more mass, while high diffusion rates release
the cosmic rays before they can couple to the gas effectively. 
Our central value 
corresponds to $\kappa_{||}=1\times10^{28}$\,cm$^{2}$\,s$^{-1}$ according
to \citet{Ruszkowski2016}. 

SNe are thought to
convert a fraction $f_{CR} = 0.1$ to~$\gtrsim$0.5 of their kinetic energy into CRs by diffusive shock
acceleration \citep[see review by][]{Helder2012}. We choose to use
$f_{CR}=0.3$  to be consistent with
\citet{SalemBryan2014}, noting that this is larger than the value of
$f_{CR}=0.1$ used in most studies in this field. 

For runs including CRs, an initial CR energy density $\eta_{CR} $ is assigned with a
value proportional to the initial gas density in each cell
$\eta_{CR}=\alpha_{CR}\times\rho$  to mimic the CR energy density
distribution of the solar neighborhood at the midplane of
$\eta_{CR}\sim3\times10^{-12}$\,erg\,cm$^{-3}$ \citep{SalemBryan2014},
where $\alpha_{CR}=10^{7}$.
 The CR energy density in starburst
galaxies is likely much higher \citep{Veritas2009, Papadopoulos2010}, but the initial distribution has a negligible
effect on the results, as CRs generated by SNe quickly diffuse and dominate the
CR dynamics. 

\subsection{Disk Model}
We use disk and star formation models for an ultraluminous infrared
galaxy (ULIRG) from \citet{Fujita2009}.
This model has a disk gas mass, $M_{g}=10^{10}$~M$_{\odot}$, a
surface density $\Sigma_{0}=10^{4}$~M$_{\odot}$\,pc$^{-2}$, and 
an exponential scale radius $R_{d}=0.7$\,kpc. 
The gravitational potential includes a disk potential based on the
thin disk approximation 
 \citep[]{Toomre1963}, and a \citet{NFW} halo potential from a halo with mass
$M_{h}=5\times10^{12}$~M$_{\odot}$, virial radius
$R_{v}=326$\,kpc, and halo concentration
factor $c=5$. We assume the gas is supported by turbulence
 with a high velocity dispersion  $\sigma =55$\,km\,s$^{-1}$, comparable
 to that observed in the molecular gas in Arp220
 \citep[]{Scoville1997}.

Only half the disk above its midplane is simulated, in an
initial grid whose 
size varies with the finest resolution employed. This was necessary to save computation
time, because the computation of CR diffusion requires time steps that
are proportional to the square of the finest resolution size
\begin{equation}
    \Delta t_{CR}<\frac{1}{6}\frac{\Delta x^2}{\kappa_{CR}}.
	\label{eq:diffusion}
\end{equation}

\subsection{Cooling}

The radiative cooling curve of \citet{SarazinWhite1987}, with a
temperature floor 
         $T_{min} = 10^{4}$\,K
is used, based on the assumption 
that the gas is kept photoionized by UV radiation from
massive stars. 
This temperature floor has a negligible effect on the dynamics
because cooling in the swept-up shells is limited by numerical
resolution, rather than by radiative cooling \citep[]{Fujita2009}. 

We note that the CR cooling losses due to Coulomb and catastrophic interactions that
we neglect could be significant in our dense ULIRG disk: for example,
the cooling time due to catastrophic losses alone is rather short, only
a few years at
$\rho\sim2\times10^{-20}$\,g\,cm$^{-3}$\citep[]{Jublegas2008}. We do not
model the CR spectrum that would be required for the
realistic cooling calculation.
The CR cooling time, both Coulomb and catastrophic,
is inversely proportional to the gas density, 
so CR cooling
losses become negligible once the CRs diffuse out of the dense disk to
its low-density outskirts and the halo beyond.  

\subsection{Starburst}
We assume a single starburst that occurs at the center of the disk.
All the kinetic energy of the starburst SNe is released
in a central wind of constant mechanical luminosity,  $L_{mech}
=10^{43}$\,erg\,s$^{-1}$. 
    This approximation can be made 
because the discrete energy inputs from
SNe generate blastwaves that become subsonic in the hot
interior of the bubble first produced by stellar winds
\citep{ML1988, Fujita2009}. This assumption means that a single superbubble forms,
evolving to produce a bipolar outflow of gas. We chose  $L_{mech}
=10^{43}$\,erg\,s$^{-1}$ as it 
corresponds to the mechanical luminosity expected
during the first 2~Myr after the onset of an instantaneous starburst
with  $M_{*}=10^{9}$~M$_{\odot}$ or of continous star formation with
100~M$_{\odot}$\,yr$^{-1}$ at $\sim$~2~Myr or 500~M$_{\odot}$\,yr$^{-1}$
at $\sim$~4~Myr
\citep{Leitherer1999}. 
Energy input is continued for the entire duration of our
simulations in this study, 
    as our simulation time 
is much less than the lifetime of the
smallest B star to go supernova.
Bubbles in our highly stratified, ULIRG disk blow
out within 1~Myr.  We note that it is an
oversimplification to assume a singe starburst site at the disk center, as
some ULIRG winds appear to require starburst regions extended to
$>1$\,kpc  \citep[]{Martin2006}. Modeling multiple star forming regions
remains as future work. 

To drive a constant luminosity wind, we add mass and energy to a
source region \citep[]{MF99} with a radius of five zones for the runs with a finest
resolution of 6.4~pc and with a radius of 5~pc for the runs with finest
resolutions of 0.8~pc and 0.2~pc. The fiducial mass input rate is
$\dot{M}_{in}=4.7$~M$_{\odot}$\,yr$^{-1}$ 
to be consistent with \citet{Fujita2009}. However, as we show later,
the mass in outflows is dominated by this injected mass, $\dot{M}_{in}$, so
we
    also 
ran the simulations with a mass input rate,
$\dot{M}_{in}=0.47$~M$_{\odot}$\,yr$^{-1}$, an order of 
magnitude lower than the fiducial value. The corresponding
temperatures of the hot, shocked, bubble interior gas are
$T_{in}=1.4\times10^{8}$ and $1.4\times10^{9}$\,K respectively. 
Radiative cooling is turned off in
the hot bubble interior with a 
   steepened 
tracer field $c$ \citep{yabe1993}.
The field
\begin{equation}
    f(c) = \tan[0.99\pi(c-0.5)].
	\label{eq:colorfield}
\end{equation}
is advected, accumulating diffusion errors that smooth
out sharp interfaces, but the inverse field $c(f)$ is factored into
the cooling function, maintaining sharp transition interfaces \citep{MF99,Fujita2004}.
This prevents 
mass numerically diffused from 
the cold, dense shell into the rarefied, hot interior from spuriously
cooling the interior, as the cooling 
time of the interior is much longer than the dynamical time of our
bubble \citep[]{ML1988}. 

\subsection{Models}
Table~\ref{tab:table1} lists the diffusion coefficients and numerical
resolutions of our models.  For comparison, 
\citet{SalemBryan2014} used a resolution of
$\Delta x_{min}=$\,61\,pc. Because of the diffusive time step
condition (Eq.~\ref{eq:diffusion}), the computation becomes
quadratically more expensive with finer resolution.  To reproduce
\citet{Fujita2009} would require
$\Delta x_{min}=$\,0.2\,pc
 in a  ($0.2$\,kpc)$^{3}$ box, 
which is impractical to do with CRs using available
resources. We had to reduce the resolution to
6.4\,pc
in the disk and 12.8\,pc outside the disk
 in a  ($3.3$\,kpc)$^{3}$ box,  
in order to study the
effects of CR pressure in the halo out to a few kiloparsecs above the
disk.

We compare the mass-loading rates of outflows driven only by SN
thermal pressure (T) and driven both by SN thermal pressure and
CR pressure (C) in models Q, W, and X with varying resolutions. 
In the highest resolution model F09, which reproduces the 2D model of
\citet{Fujita2009} in 3D, we run a simulation of an outflow driven by SN
thermal pressure alone.
We also run the lowest resolution model with a low mass injection rate, Z, 
and we test the effects of varying diffusion coefficients using the
lowest resolution models, ZI and ZC. 

\begin{table}
	\caption{Numerical models 
        }
	\label{tab:table1}
	\begin{tabular}{llcccrl} 
		\hline
		Model & Physics$^{a}$ & $\Delta x_{min}$ (pc) & base &
                                                                       AMR$^{b}$ &   $\kappa_{CR}^{c}$  & $\dot{M}_{in}^{d}$ \\
		\hline
		F09 & T     & 0.2        & $64^3$              & 6 &
                                                                     --
                                                                                                        & 4.7
                                                                                                      
          \\  
		Q    & T/C & 0.8        & $64^3$                & 4 &
                                                                      10 & 4.7\\
                W    & T/C & 3.2        & $16^2\times32$ & 3 & 10 &
                                                                    4.7\\
                X    & T/C & 6.4/12.8         & $16^3$ & 4 & 10 &
                                                                    4.7\\
		Z    & T/C & 6.4/12.8 & $16^3$               & 4 & 10 & 0.47\\
                ZI  & C     & 6.4/12.8 & $16^3$               & 4 & 1 & 0.47\\
		\hline
	\end{tabular}
$^{a}${SN contribution: T = Thermal pressure; C = Cosmic rays} \\
$^{b}${Number of adaptive mesh refinement levels} \\
$^{c}${In units of $3 \times 10^{26}$\,cm$^{2}$\,s$^{-1}$}\\
$^{d}${Mass input rates in units of M$_{\odot}$\,yr$^{-1}$}\\
\end{table}

\section{Results}
\begin{figure*}
\includegraphics[width=.8\columnwidth]{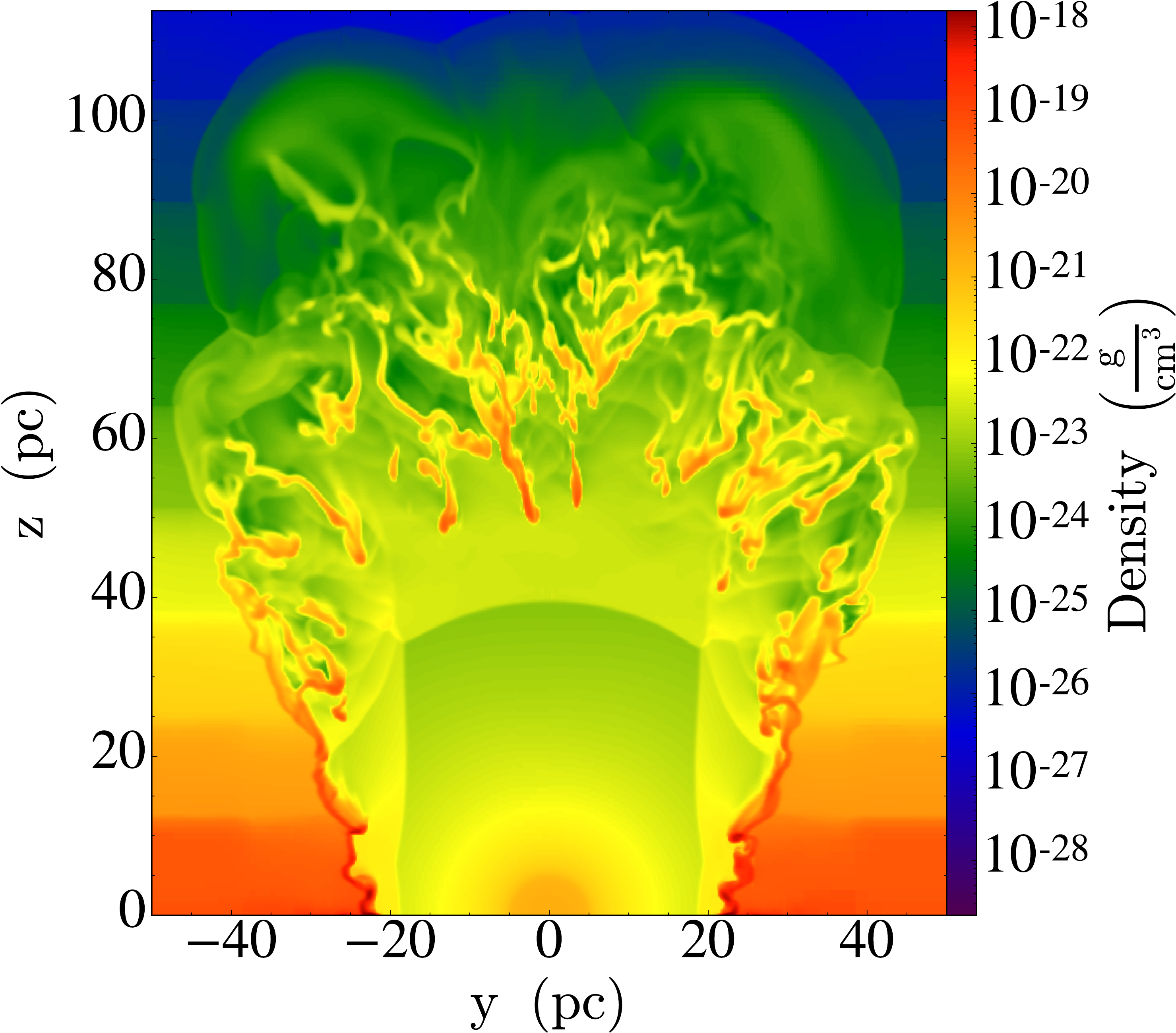}
\includegraphics[width=.8\columnwidth]{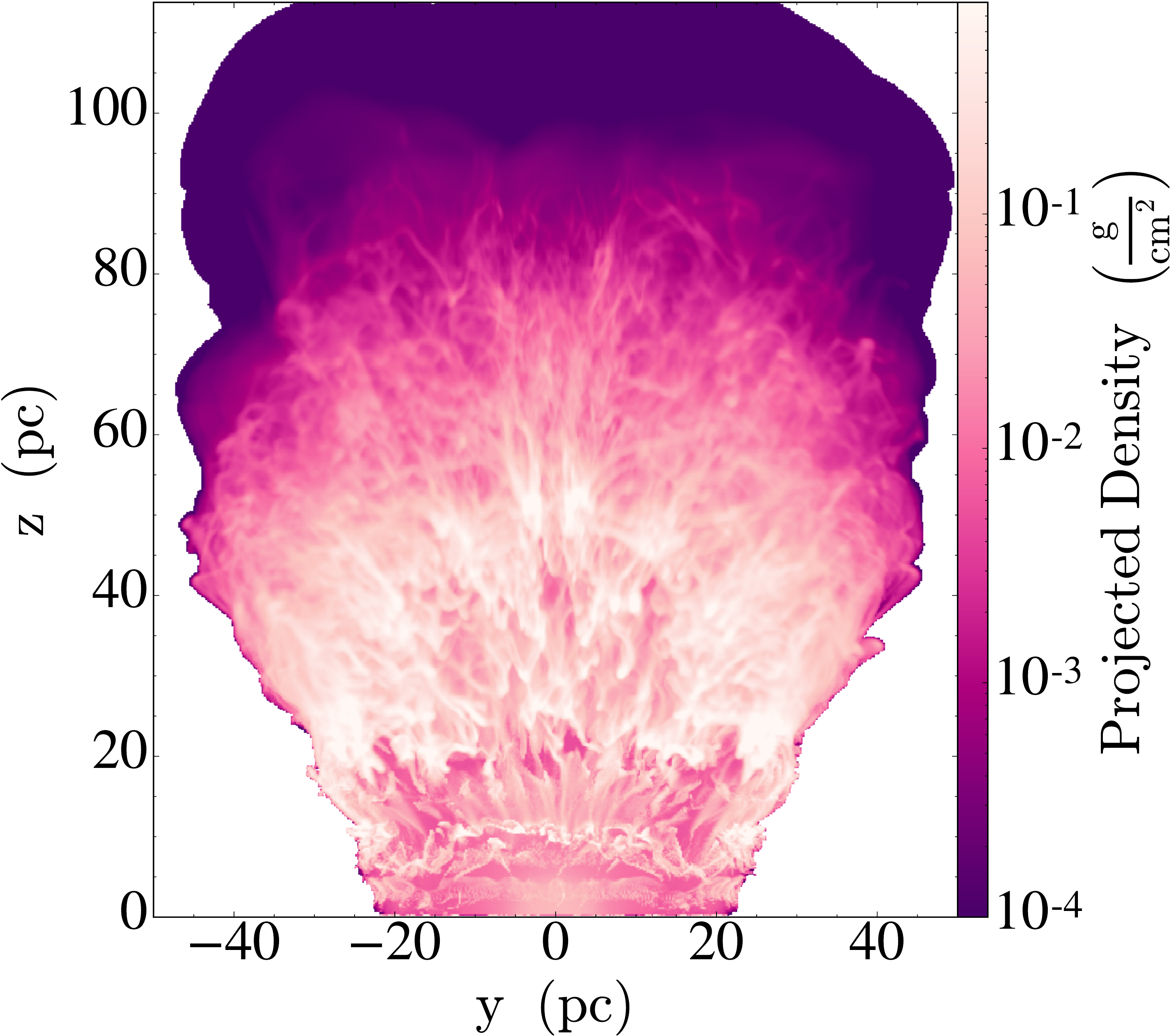}
    \caption{A density slice at the center (\textit{left}); and a
      projected density distribution (\textit{right}) of a thermal
      pressure driven outflow blowing out
      of the disk at t$\sim0.16$\,Myr in our highest resolution run
      F09, with $\Delta x_{min}=0.2$\,pc. The fine resolution in 3D
      resolves the fragmentation of the swept-up shells into spikes by
      Rayleigh-Taylor instability.  
We confirm that our 3D result is consistent
    with our previous 2D result \citep[]{Fujita2009}  in the
    kinematic behaviour of the hot interior and the cooler shells. Only gas
    moving at $v_{z}>0$ is used for the projection,
  to exclude quiescent background gas in the disk and halo.  Note that the color scales used are the same as in
     Figure~~\ref{fig:lowdensitybubble} and Figure~\ref{fig:big} below.}
    \label{fig:y39}
\end{figure*}
\begin{figure*}
\includegraphics[width=.8\columnwidth]{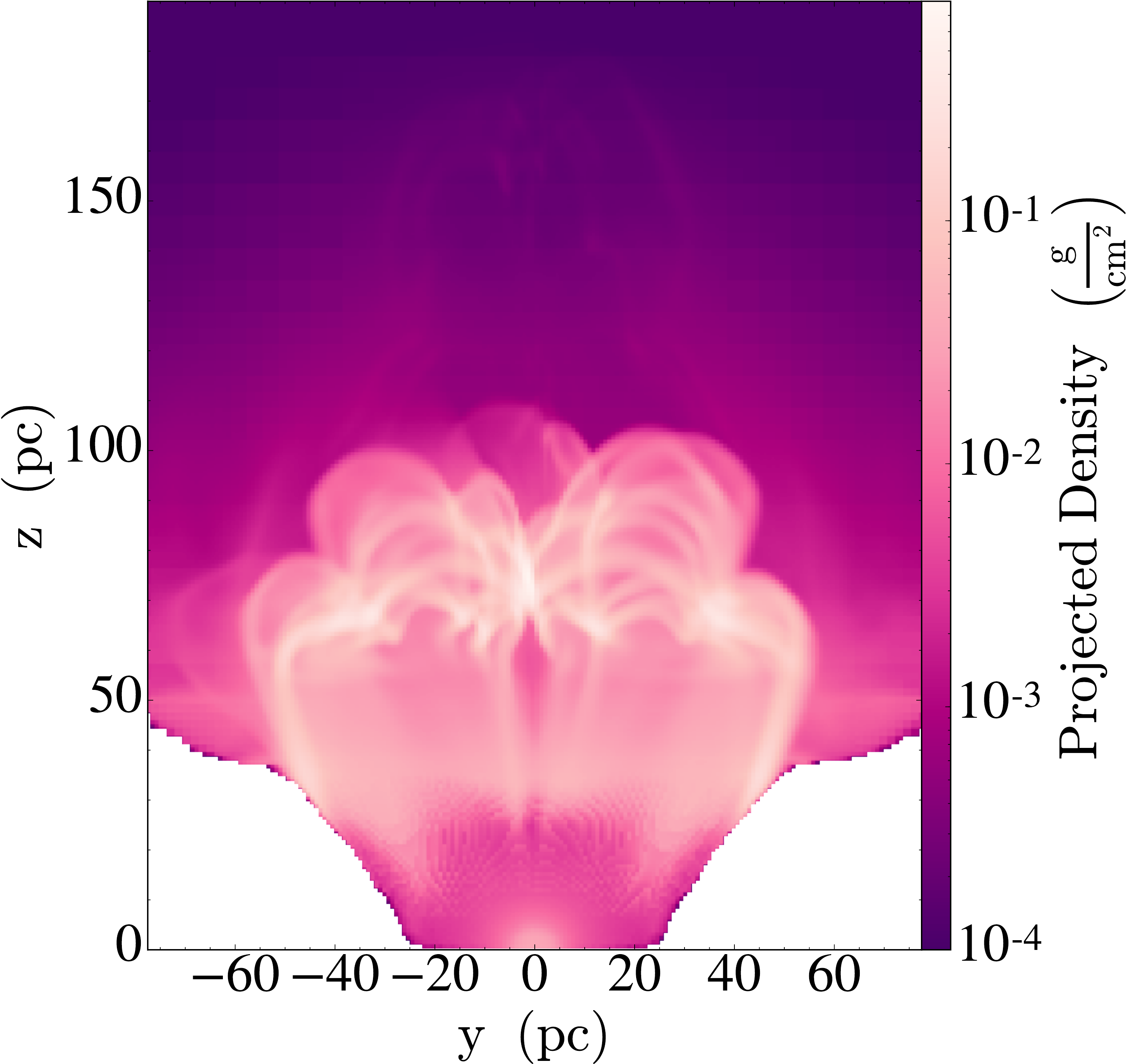}
\includegraphics[width=.8\columnwidth]{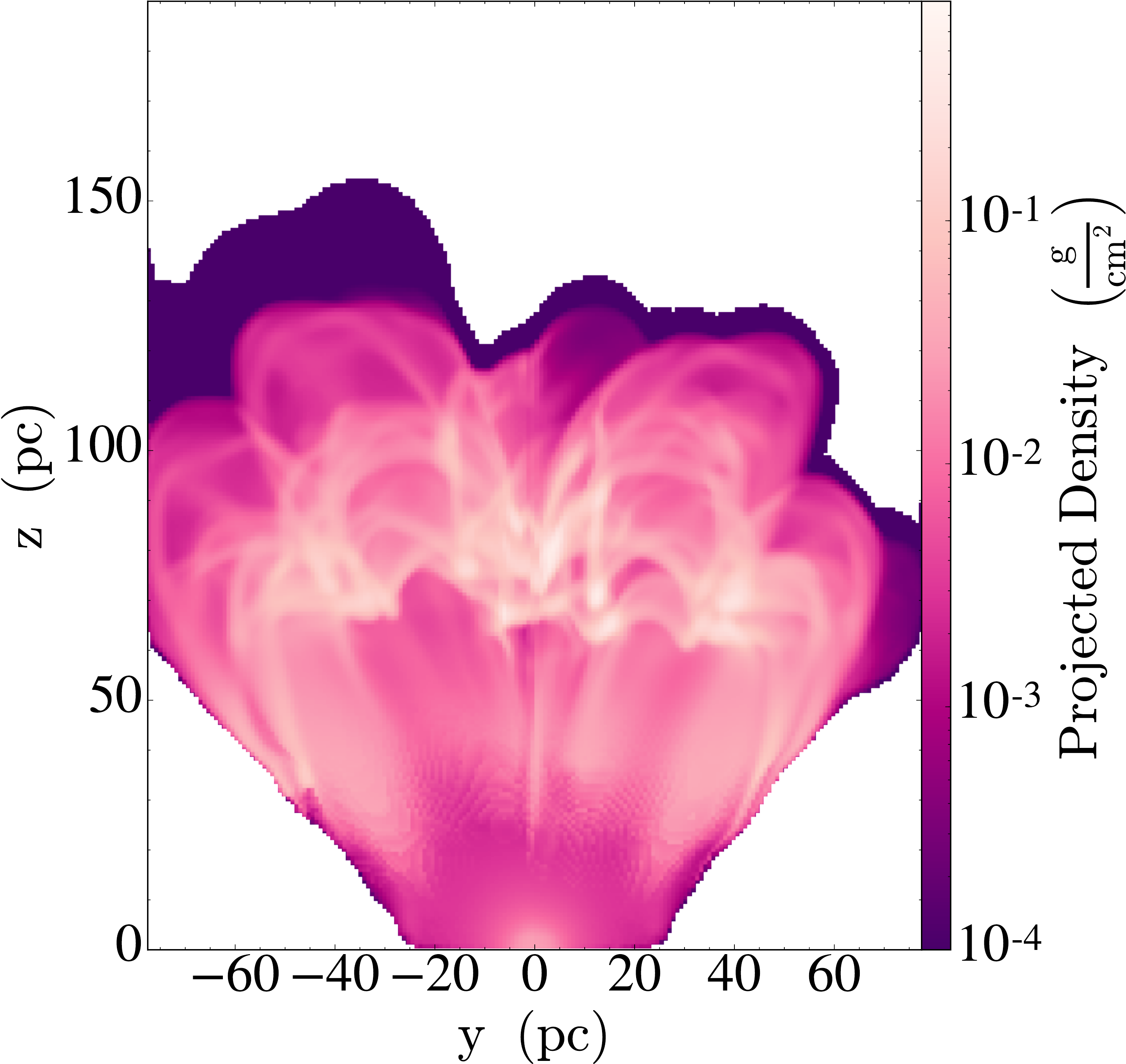}
    \caption{
      Projected density distributions of a CR pressure driven outflow
      (QC; {\em left}) and a thermal pressure driven outflow (QT; {\em
        right}) blowing out 
      of the disk at $t\sim0.27$\,Myr in our highest resolution
      run with $\Delta x_{min}=0.8$\,pc. At this time, the thermally driven shock has
      grown faster, but CRs diffusing out from the CR driven bubble have already
      accelerated the halo gas above. 
Again only the gas
    moving at $v_{z}>0$ is used for the projected density distribution.
}
    \label{fig:q1}
\end{figure*}

Figure~\ref{fig:y39} shows our 3D version F09 of the fiducial 2D model by \citet{Fujita2009}  of
a galactic outflow driven by SN thermal pressure alone with
$\Delta x_{min}=$\,0.2\,pc 
using 6 levels of refinement. The kinematic behaviour of the outflow
agrees between the previous 2D simulation and our 3D
version. The shocked swept-up shells of ambient gas fragment by
Rayleigh-Taylor (RT) instability, showing the usual spike and bubble
morphology. Most fragments travel at a few
hundred kilometers per second, while a small fraction of them exceed
the escape velocity $v_{esc}=800$\,km\,s$^{-1}$ at
$0.01R_{v}$.

We do find that the 3D computation resolves the growth of RT spikes,
while in 2D, the assumption of azimuthal symmetry limited RT instabilities to grow
as rings, decreasing the extent of fragmentation. Figure~\ref{fig:y39} 
shows the hot, low-density interior gas
streaming through the spikes, ablating their outer layers by Kelvin-Helmholtz
instability. 

Figure~\ref{fig:q1} compares projected density distributions of a
superbubble driven by SN thermal pressure alone (QT) and a superbubble
driven by both SN
thermal and CR pressure (QC), blowing out of the disk at $t=0.27$\,Myr
in runs with  $\Delta x_{min}=$\,0.8\,pc. Only gas rising vertically out of the
disk at $v_{z}>0$\,km\,s$^{-1}$ is projected in the figure to
exclude the quiescent disk and halo gas. The
classic thermal pressure driven superbubble (QT) is larger, with the
shell fragmenting, and letting the hot gas escape, compared to the CR
driven superbubble with CRs quickly diffusing
into the disk beyond the swept-up shells and
 out into the halo.
\begin{figure*}
\includegraphics[width=.68\columnwidth]{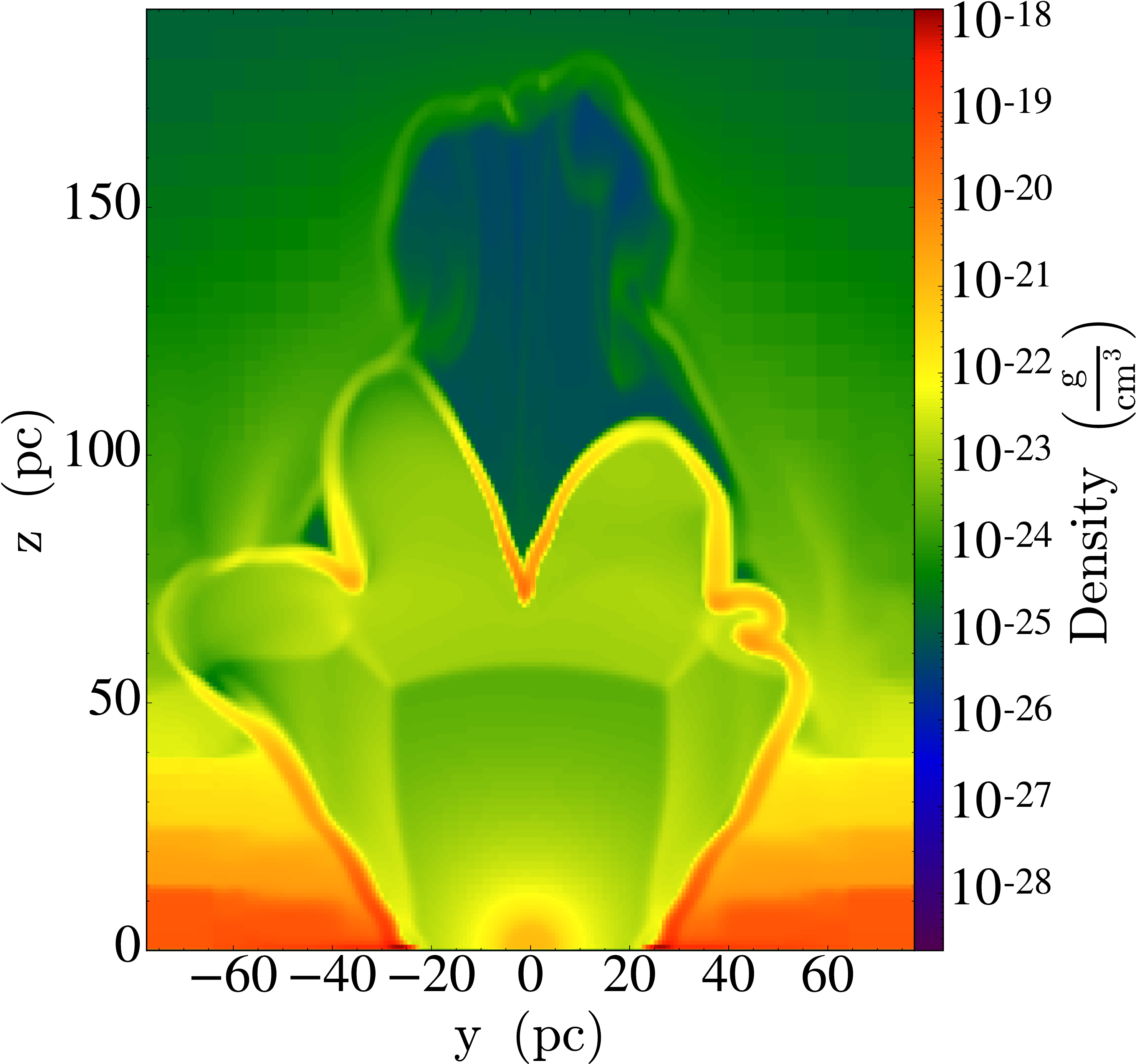}
\includegraphics[width=.68\columnwidth]{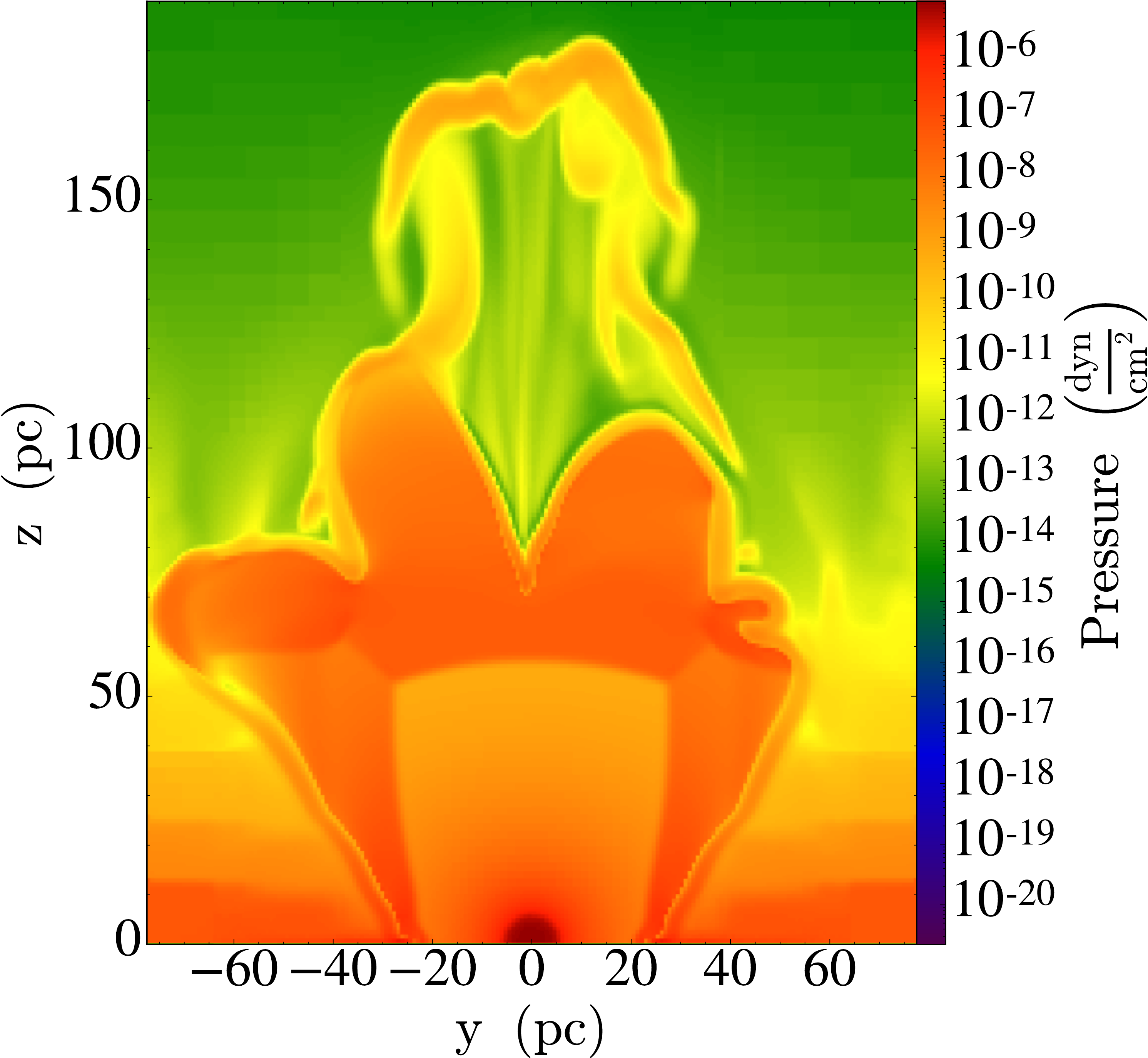}
\includegraphics[width=.68\columnwidth]{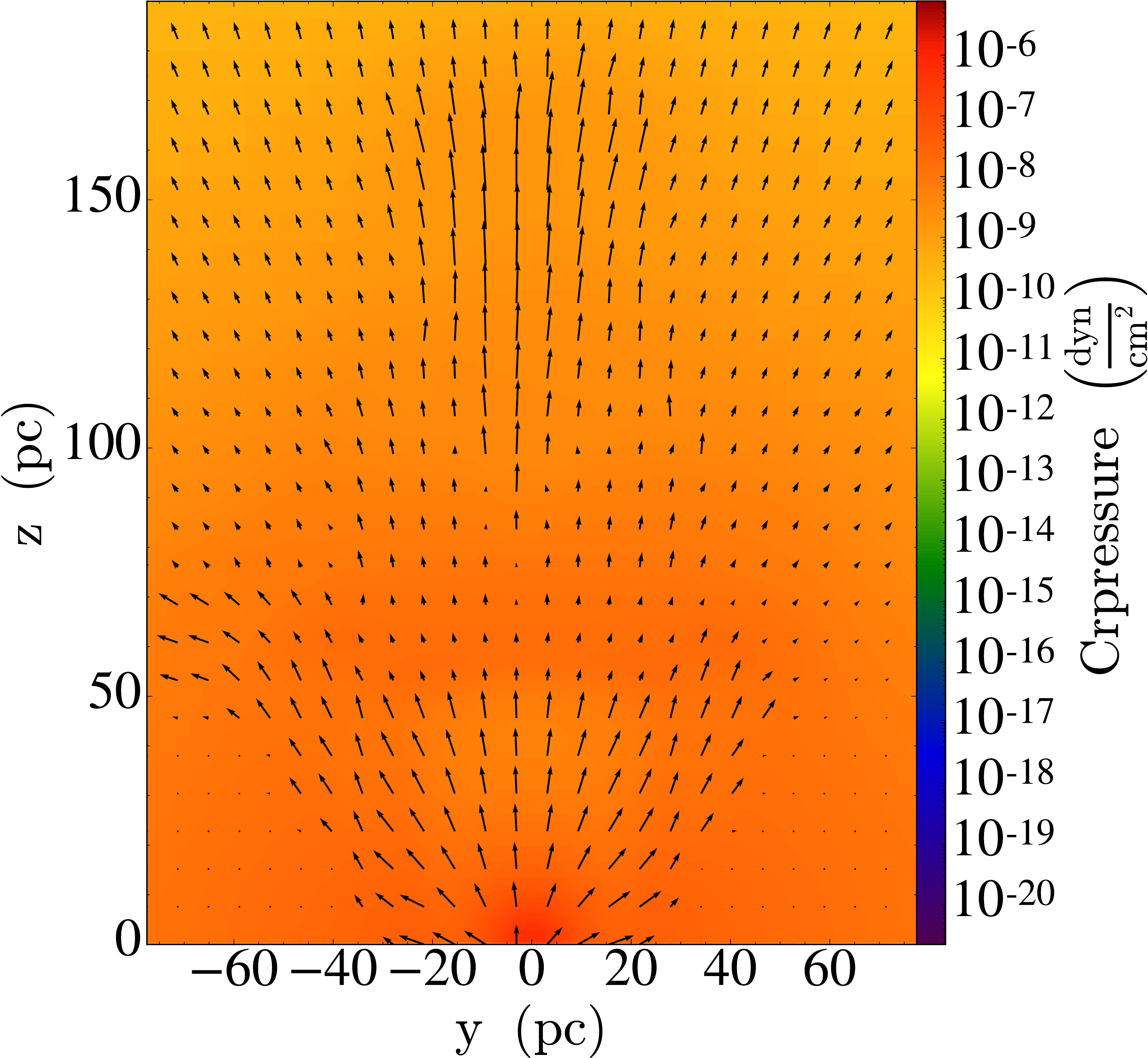}
    \caption{
      Slices of density (\textit{left}), pressure (\textit{middle}), 
      and CR 
     pressure overlaid with velocity (\textit{right}) in the plane in a CR
      pressure driven outflow at $t\sim0.27$\,Myr in model QC. The CR pressure
      gradient across the outer edges of the swept-up shell drives a
      low density bubble that expands into the halo at
      $v>1000$\,km\,s$^{-1}$ ahead of the
      thermal pressure driven superbubble. 
   The maximum velocity shown is $v_{max}=3700$\,km\,s$^{-1}$. 
     Note that the color scales used are the same as in
     Figure~~\ref{fig:y39} and Figure~\ref{fig:big} below.
    \label{fig:lowdensitybubble}
}
\end{figure*}
\begin{figure}
	\includegraphics[width=\columnwidth]{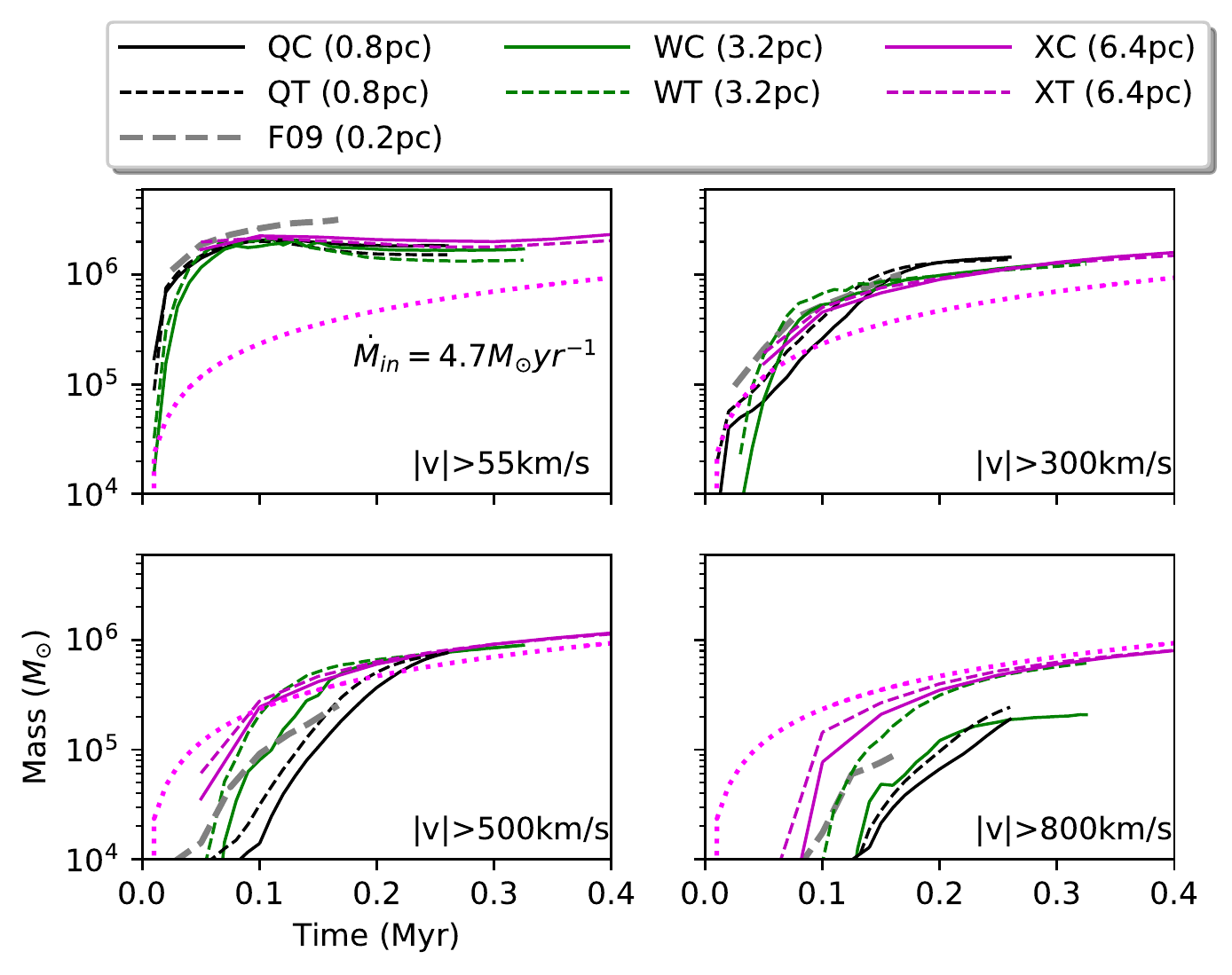}
    \caption{ 
 Total mass 
    with (\textit{solid}) and without (\textit{dashed}) CR pressure 
in outflows moving upward ($v_{z}>10$\,km \,s$^{-1}$) with 
     speeds 
      $|\bm{v}|>55$\,km \,s$^{-1}$, $|\bm{v}|>300$\,km
      \,s$^{-1}$, $|\bm{v}|>500$\,km \,s$^{-1}$, and
      $|\bm{v}|>800$\,km \,s$^{-1}$
      in our models with a mass injection rate
      $\dot{M}_{in}=4.7$~M$_{\odot}$\,yr$^{-1}$ 
    (cumulative input mass shown in {\em 
   magenta 
dotted})
at varying resolutions
    given in the legend.
We only have results up to $t\sim 0.15$~Myr for the highest
resolution run (\textit{dashed gray}) due to limited computational
resources, but the total mass moving at $|\bm{v}|>55$\,km \,s$^{-1}$ seems to
have already
reached a plateau. 
Only a small fraction of mass is moving out into the
    halo. The total mass in outflows with varying resolutions
  converges well after $\sim$ 0.2--0.3~Myr, but it is dominated by the
  SN ejecta mass $\dot{M}_{in}$ that
  generates the outflows, 
   rather than by entrained gas.  
    \label{fig:massloss-multiple1}
}
\end{figure}
\begin{figure}
	\includegraphics[width=\columnwidth]{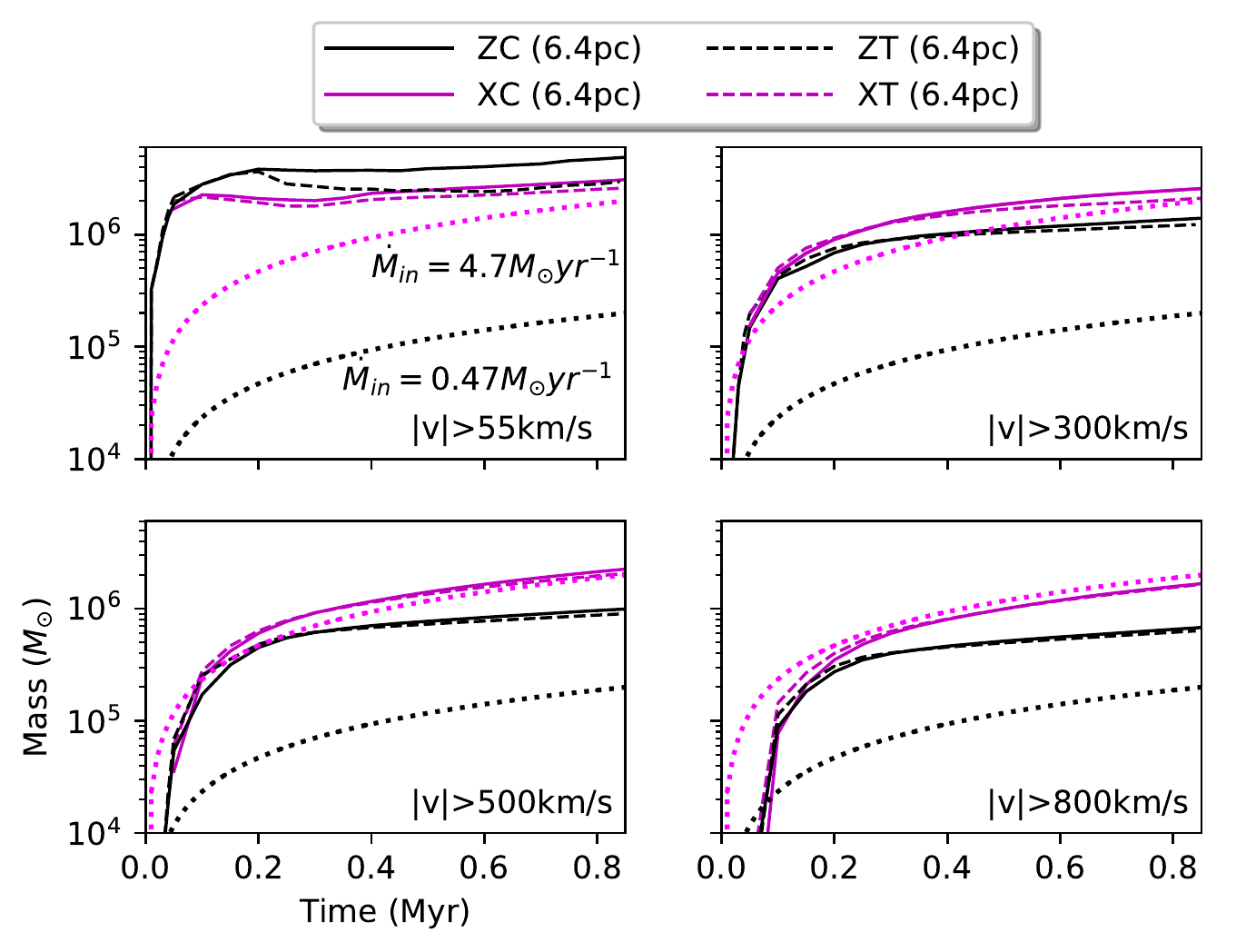}
    \caption{
Total mass
    with (\textit{solid}) and without (\textit{dashed}) CR pressure  
in outflows moving upward ($v_{z}>10$\,km \,s$^{-1}$) with 
     speeds 
      $|\bm{v}|>55$\,km \,s$^{-1}$, $|\bm{v}|>300$\,km
      \,s$^{-1}$, $|\bm{v}|>500$\,km \,s$^{-1}$, and
      $|\bm{v}|>800$\,km \,s$^{-1}$
 in our lowest resolution models
      ($\Delta x=6.4/12.8$~pc) 
    with mass injection rates, $\dot{M}_{in}=4.7$~M$_{\odot}$\,yr$^{-1}$ 
    (cumulative mass injected in 
{\em magenta dotted})  and $\dot{M}_{in}=0.47$~M$_{\odot}$\,yr$^{-1}$ 
         ({\em black dotted}).
The amount of mass carried in the outflows is approximately the
same independent of the mass input that we use to generate the outflows. 
    \label{fig:massloss-multiple2}
}
\end{figure}

The resulting CR
pressure gradient accelerates the low-density halo gas to
$v>500$\,km\,s$^{-1}$, and in particular, drives a low-density bubble 
    beyond
the outer edge of the swept-up shells that travels at
$v>1000$\,km\,s$^{-1}$, as shown in
Figure~\ref{fig:lowdensitybubble}. This bubble
initially expands faster into the halo than the thermal pressure
driven shock, but it only carries $\sim10\%$ of the total
mass traveling at $v>1000$\,km\,s$^{-1}$ on the grid
in model QC. The total mass moving 
upwards with vertical velocity  $v_{z}>10$\,km\,s$^{-1}$ and
speed exceeding the turbulent sound speed $|{\bf v}|> c_s = 55$\,km\,s$^{-1}$ 
is $1.5\times10^{6}$~M$_{\odot}$ in the
thermal-pressure driven
case and $1.8\times10^{6}$~M$_{\odot}$ in the CR driven case, 
     only a modest increase.  
For the mass-loading calculation, we exclude mass moving at
$v_{z}\le10$\,km\,s$^{-1}$ in order to only include gas accelerated by
bubbles. Note that the gas moving with $v_{z}=10$\,km\,s$^{-1}$  will 
travel only $\sim$~5~kpc upward in 500~Myr if it maintains its speed.

In order to study the effects of CR pressure accelerating mass, we ran the same simulations on
larger grids covering several kiloparsecs in the halo, by reducing the
resolution (Table~\ref{tab:table1}).
Figure~\ref{fig:massloss-multiple1}  shows a resolution study of the
total mass in outflows 
     modeled
with varying resolutions, with and without CR pressure. 
The mass injection rate
    for these runs 
is $\dot{M}_{in}=4.7$~M$_{\odot}$\,yr$^{-1}$.
The peaks of mass entrainment tend to
occur later in the higher-resolution runs because blowout occurs later due to
the smaller sizes of their source regions (which have constant size in
zones), but after the peaks, the total
masses converge within a factor of a few. All models seem to
converge well after $\sim$0.2--0.3~Myr, as
the outflow structures
are fully developed by this time. However, by the same time, 
the total mass of outflow gas is nearly equal to the mass we inject to
generate the outflow (\textit{dotted} in
Figure~\ref{fig:massloss-multiple1}). 

\begin{figure*}
\includegraphics[width=.67\columnwidth]{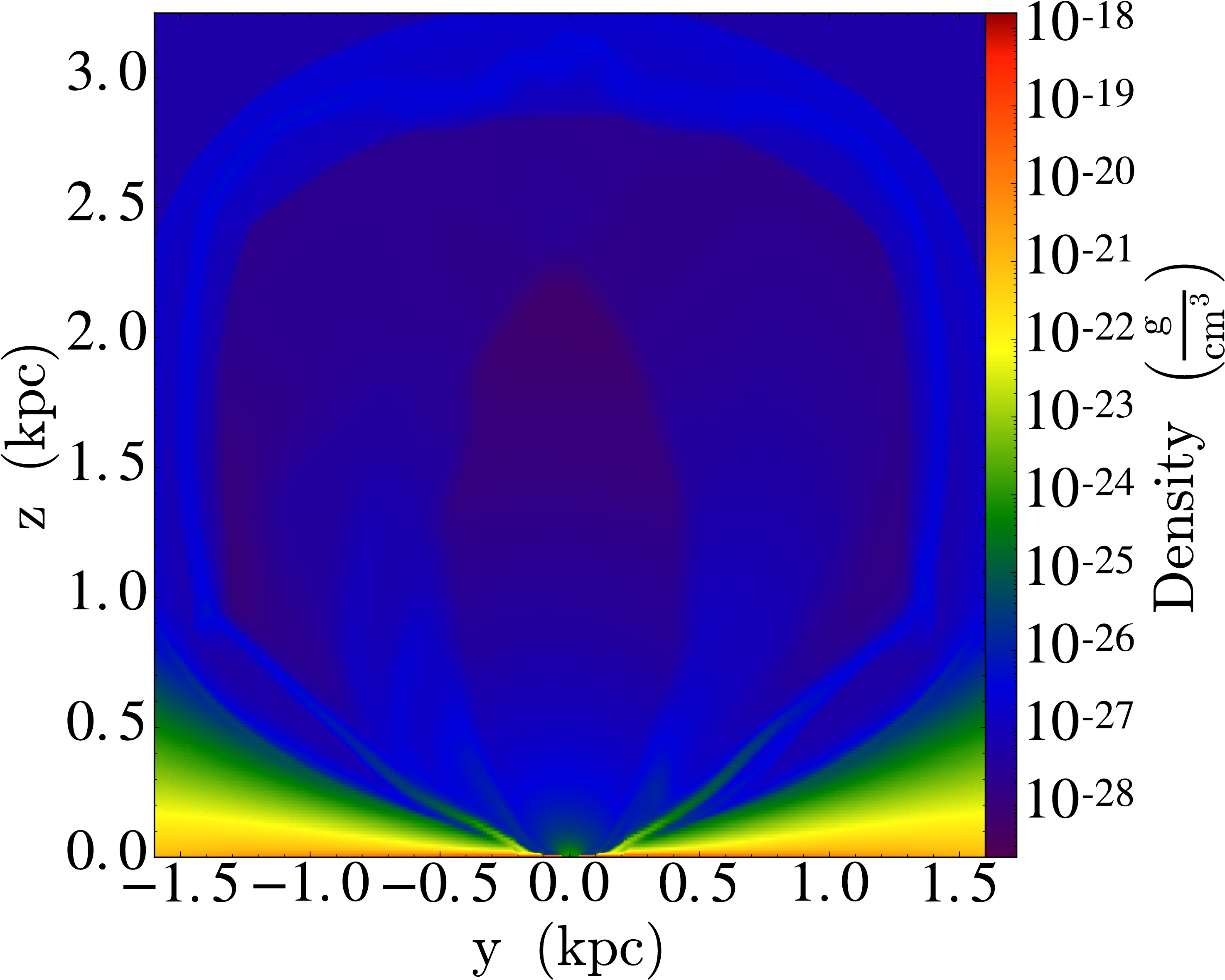}
\includegraphics[width=.67\columnwidth]{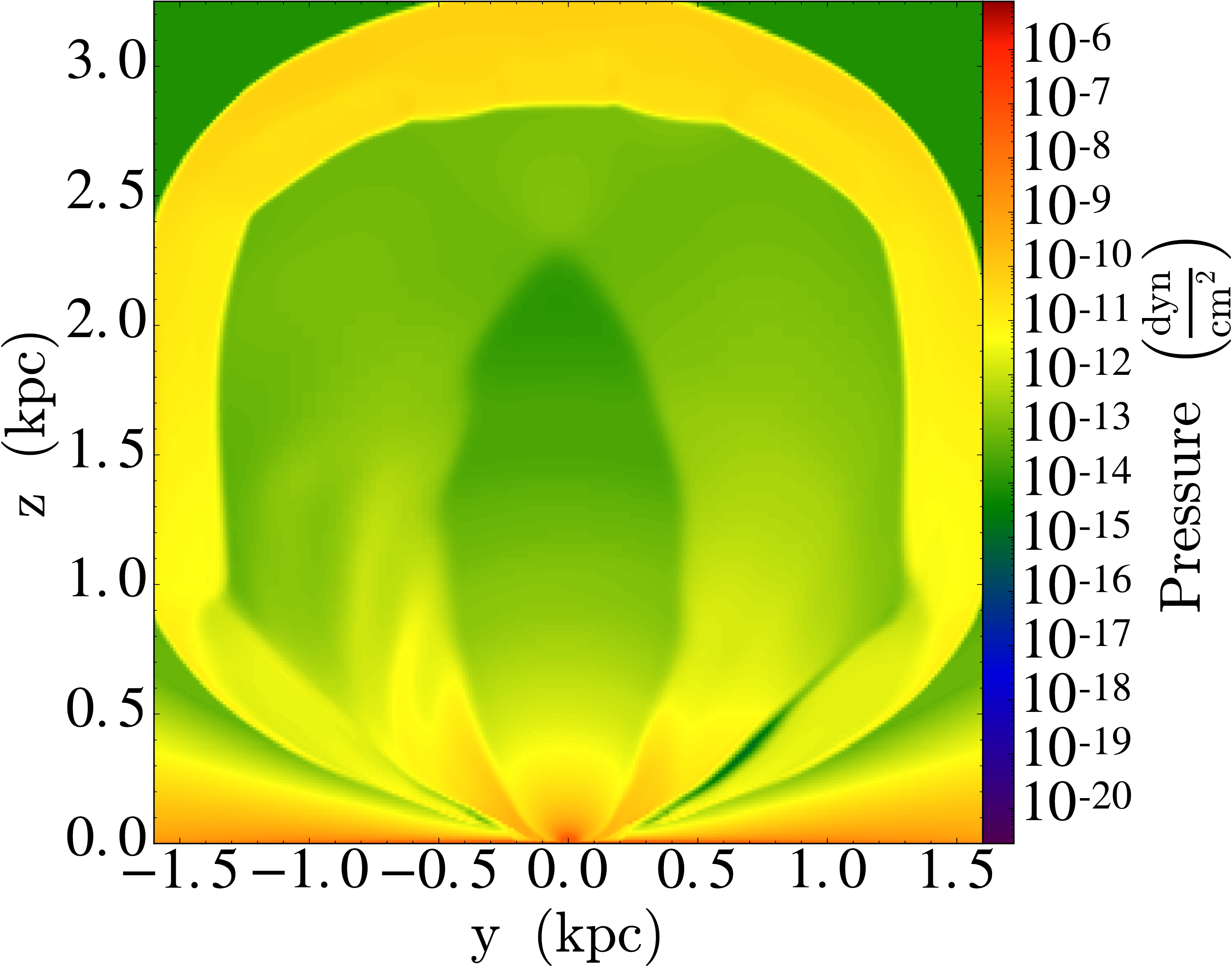}
\includegraphics[width=.67\columnwidth]{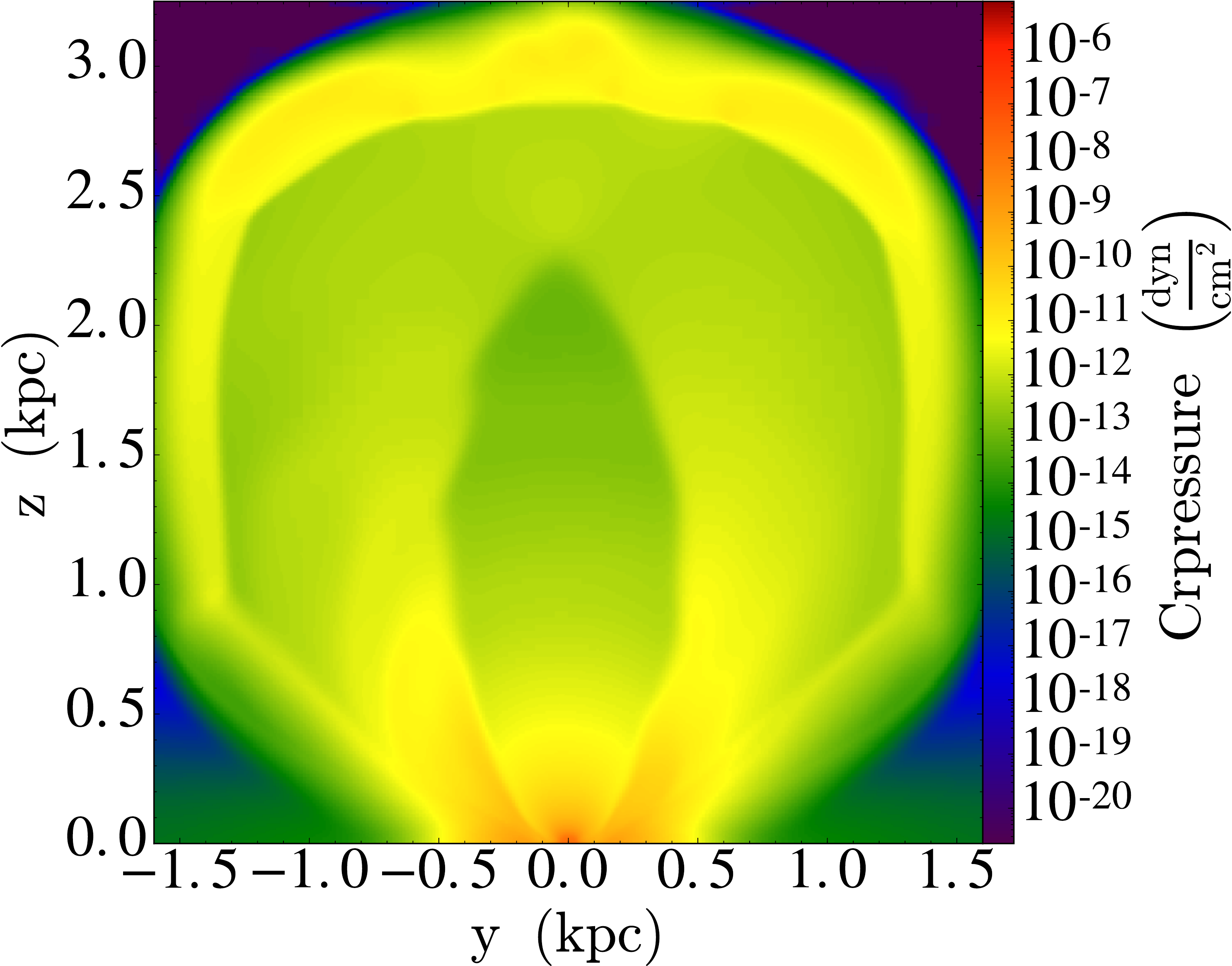} 
\includegraphics[width=.6\columnwidth]{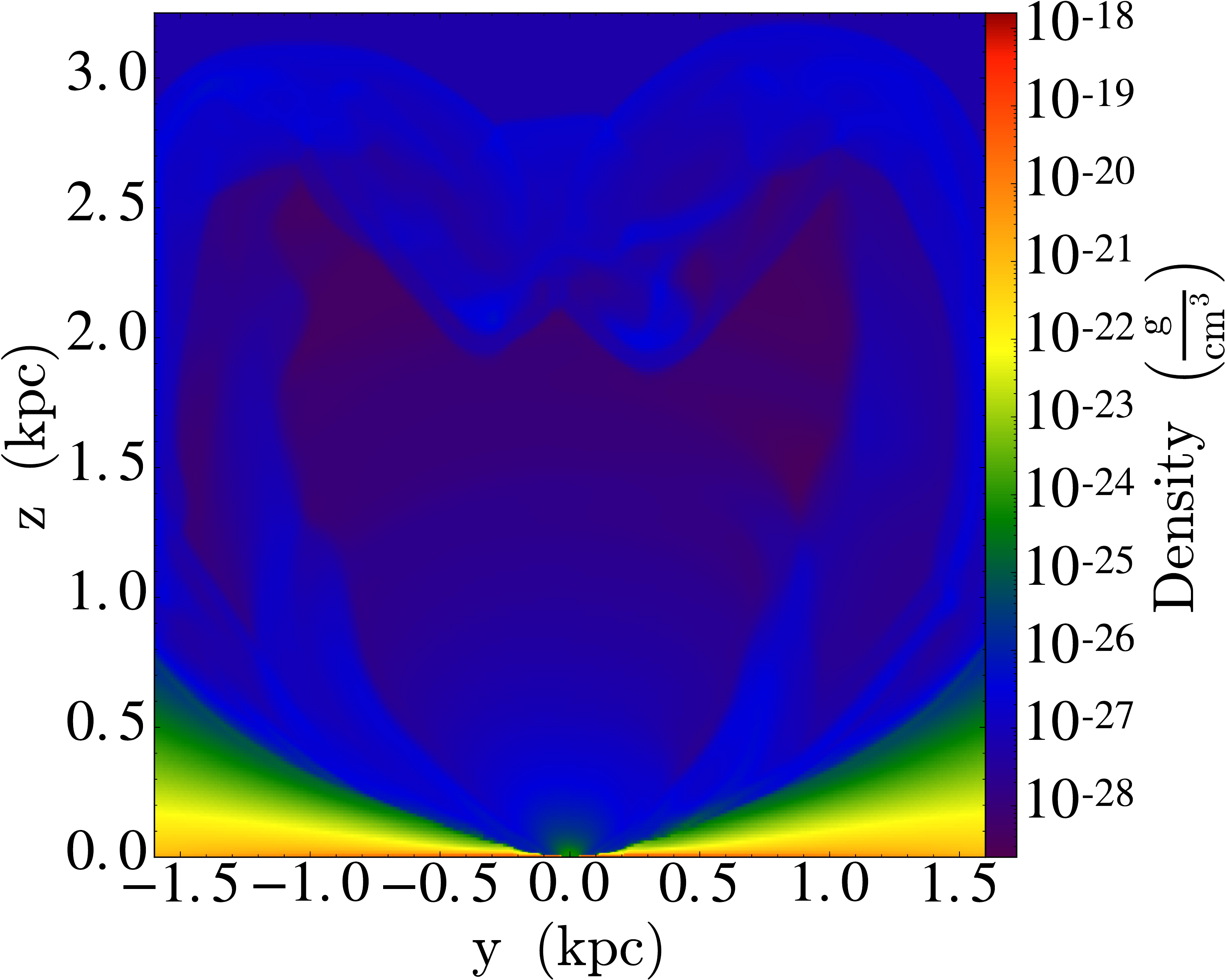}
\includegraphics[width=.6\columnwidth]{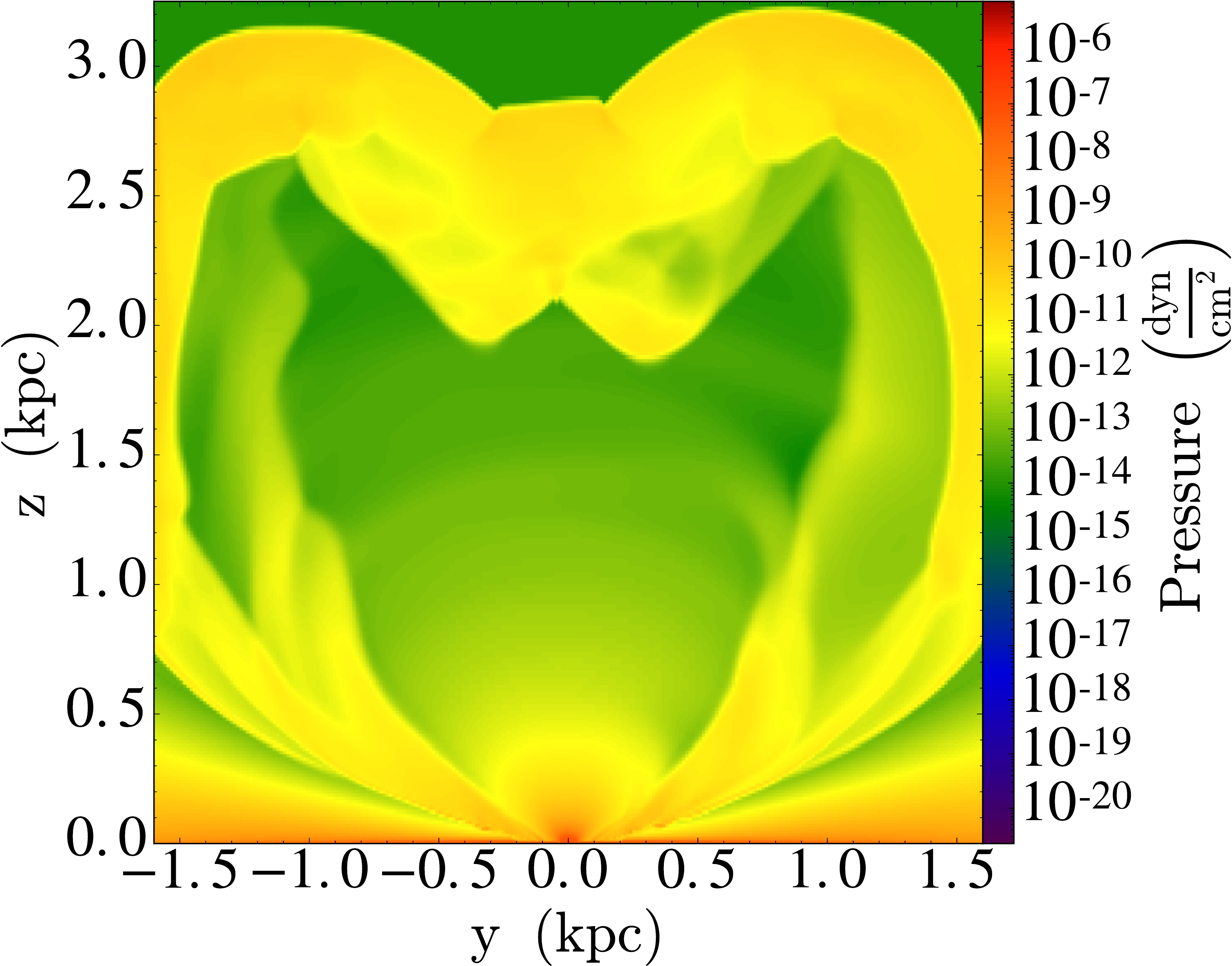}
    \caption{Slices of density (\textit{left}), pressure (\textit{middle}), and CR
      energy and gas velocity(\textit{right}) distributions in a CR
      pressure driven outflow (ZC: \textit{top}), and slices of density
      (\textit{left}) and pressure (\textit{right}) distributions in
      a thermal pressure driven outflow (ZT: \textit{bottom})
      in our lowest-resolution run ($\Delta x_{min}=6.4$~pc in the
      disk and 12.8~pc outside) that covers several kiloparsecs in the halo 
      at $t=1.2$\,Myr. The mass injection rate is
      $\dot{M}_{in}=0.47$~M$_{\odot}$\,yr$^{-1}$. 
      By this time, the thermal pressure driven shock has
      overtaken the low-density bubble created by CR pressure earlier
      in Figure~\ref{fig:lowdensitybubble}, and
 both CR pressure and thermal pressure driven outflows are behaving 
    similarly.   
     Note that the color scales used are the same as in
     Figure~\ref{fig:y39}  and Figure~\ref{fig:lowdensitybubble}.
    \label{fig:big}
}
\end{figure*}

In order to differentiate ambient gas from injected mass in
the outflows, we run simulations with a mass
injection rate an order of magnitude smaller than the fiducial
value, $\dot{M}_{in}=0.47$~M$_{\odot}$\,yr$^{-1}$, using the lowest
resolution (Z) since the resulting extremely hot interior gas lowers
the timestep substantially. 
Figure~\ref{fig:massloss-multiple2} shows $\sim~10^{6}$~M$_{\odot}$ of
disk gas is accelerated in the outflow in Z runs, and 
total outflow masses are the same within
a factor of two, whether the mass injection rate is 4.7 or
$0.47$~M$_{\odot}$\,yr$^{-1}$, with or without CR pressure. The
small difference is due to the temperature difference of the bubble interiors:
there is less cooling of numerically diffused gas from
the swept-up shells in a hotter, lighter bubble than a cooler, denser bubble.

The mass entrained is plotted in multiple
velocity ranges because previous studies suggested that the gas
accelerated by CR pressure is more diffuse, and moves at much
lower velocity than the gas accelerated by SN thermal pressure, which moves
at higher velocity
   due to its higher sound speed
\citep[]{SalemBryan2014,Simpson2016, Girichidis2016, Pakmor2016}. In our simulations, there is no
noticeable difference in mass-loading rates with and without CR
pressure, so this trend is absent. 

The convergence we find 
after the peaks in outflow masses 
allows us to use the lowest resolution runs to study the effects of CR
pressure at later times when the outflows
spread out to a few kiloparsecs in the halo.
Figure~\ref{fig:big} shows outflows in our model disk with and without
CR pressure at $t=1.2$~Myr. Our disk is very thin due to our massive
disk potential with a disk gas mass $M_{g}=10^{10}$~M$_{\odot}$ and a
surface density $\Sigma_{0}=10^{4}$~M$_{\odot}$\,pc$^{-2}$. It 
    flares
out to a few hundred parsecs 
at 
the exponential scale radius
$R_{d}=0.7$\,kpc and 
   reaches a thickness of $\sim$~1~kpc at $\sim 2 R_d$.

In simulation ZC, we find
that by $t\sim0.4$\,Myr, the thermal-pressure driven superbubble
catches up with the low-density CR driven 
bubble above it, and eventually overtakes it, and pushes it to the side.  
CRs quickly diffuse through the disk and the upper atmosphere 
outside the 
   shells swept-up 
by the superbubble, for example, 
    to a
radius of $\gtrsim$0.5~kpc by $t=1.2$~Myr.
(\textit{top right} in Figure~\ref{fig:big}). However, most of the
disk gas 
    remains
tightly bound in the galactic potential, not entrained in the outflow. As a
result, both CR pressure and thermal pressure driven outflows behave
    quite similarly.
CR pressure can accelerate $\sim60~\%$ more mass than thermal pressure alone at
vertical velocity $v_{z}>10$\,km\,s$^{-1}$, and $\sim45~\%$ at
$v_{z}=55$\,km\,s$^{-1}$. The total mass moving is only
$3.1\times10^{6}$~M$_{\odot}$ and $2.7\times10^{6}$~M$_{\odot}$
respectively, even in the CR-driven outflow. 

\begin{figure}
	\includegraphics[width=\columnwidth]{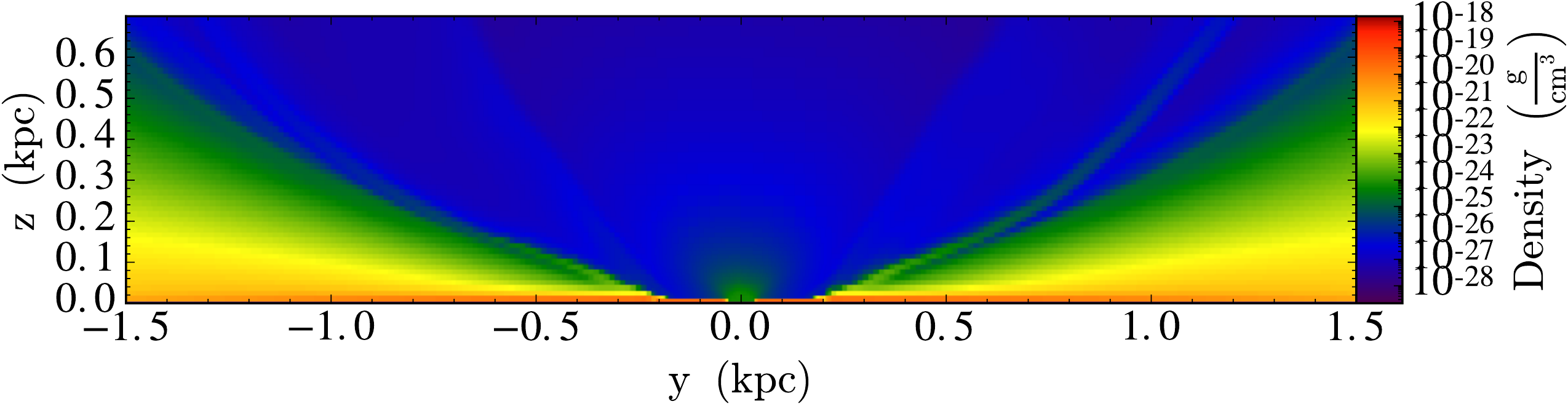}   
\includegraphics[width=\columnwidth]{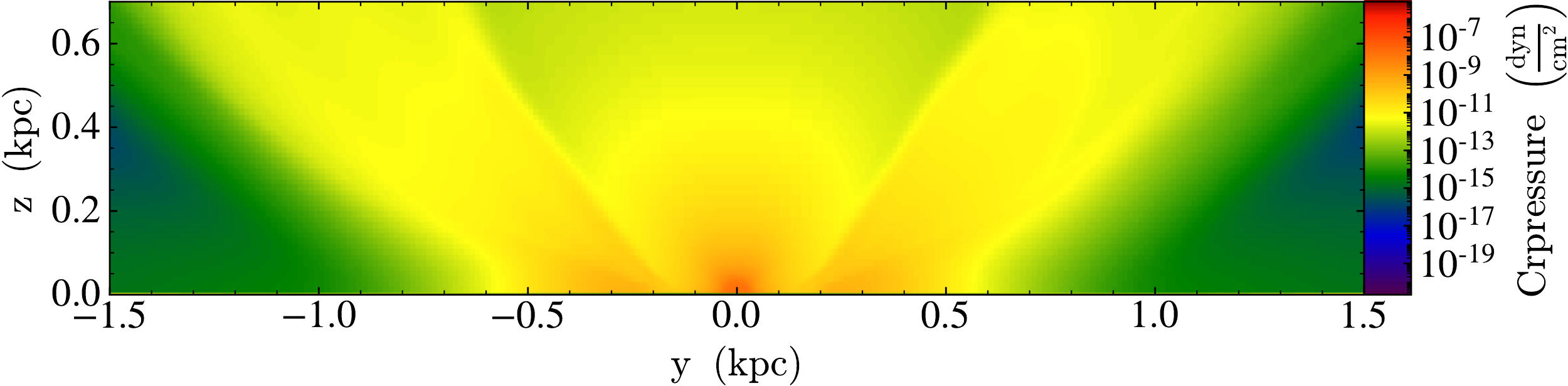}
\includegraphics[width=\columnwidth]{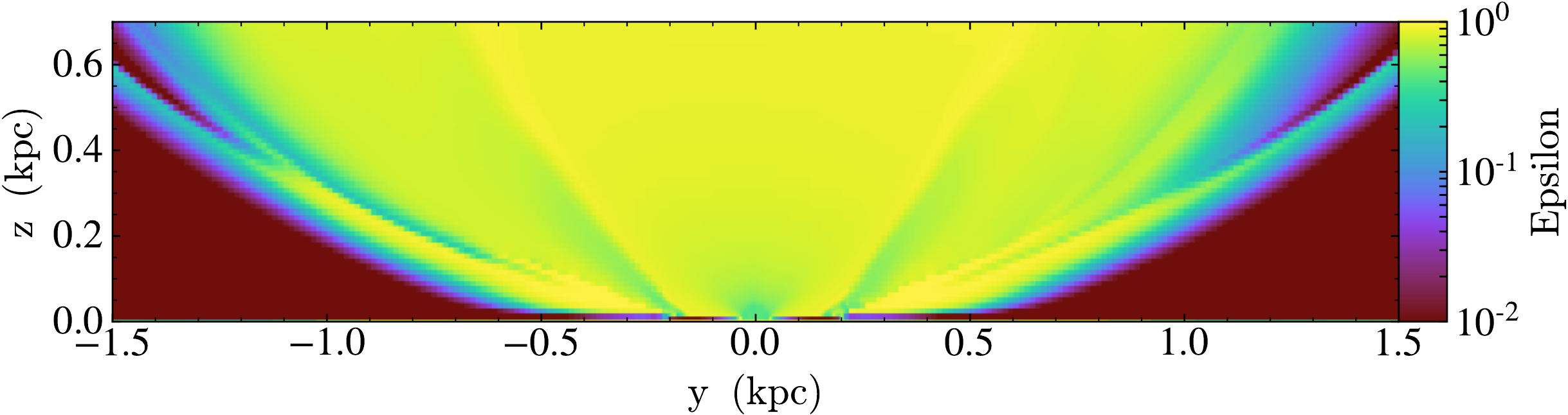}
 \caption{
Slices of density (\textit{top}), CR pressure  (\textit{middle}), and epsilon
      (\textit{bottom}) distributions 
      in the disk with a CR
      driven outflow with a diffusion coefficient
      $\kappa_{CR}=3\times10^{27}$ (ZC) 
      in our lowest-resolution runs ($\Delta x_{min}=6.4$~pc in the
      disk and 12.8~pc outside) at $t=2.2$\,Myr. The mass injection rate is
      $\dot{M}_{in}=0.47$~M$_{\odot}$\,yr$^{-1}$. CRs diffuse out to the
      halo and into the disk, dominating the pressure in the outflow,
      but are unable to entrain the disk gas. }
   \label{fig:central}
\end{figure}
\begin{figure}
	\includegraphics[width=1.1\columnwidth]{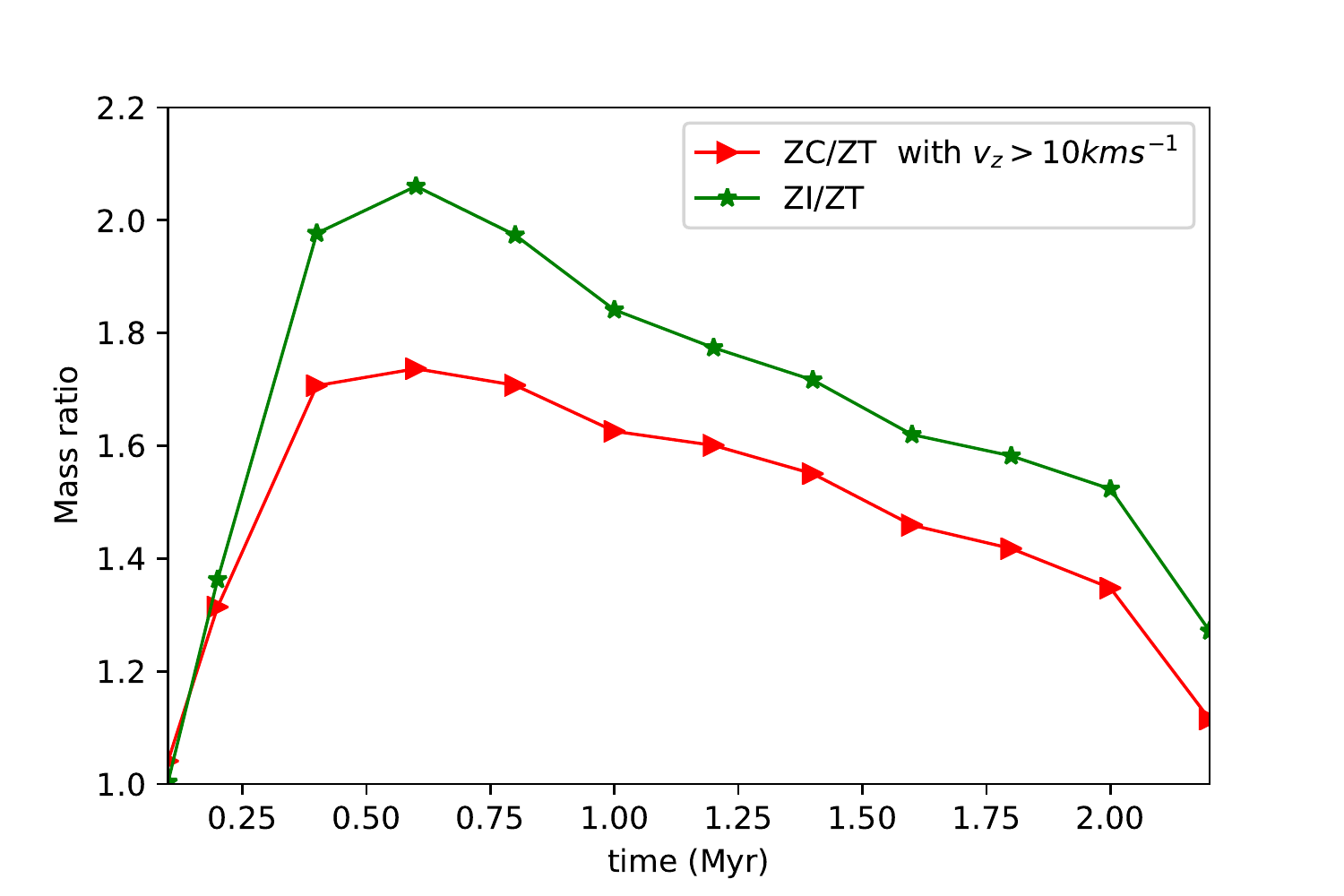}
    \caption{
The ratio of  total mass in CR-driven outflows  to total mass in
thermally-driven outflow, with CR diffusion
 coefficients $\kappa_{CR} =3\times10^{27}$  (ZC; \textit{red triangle}) and  $3\times10^{26}$\,cm$^{2}$\,s$^{-1}$
       (ZI; \textit{green star}).  Outflow mass includes all mass moving upward with
       $v_{z,min}=10$\,km\,s$^{-1}$. CR pressure 
      initially loads about twice as much mass as thermal pressure
      alone at blowout,
      but does not seem to accelerate more mass afterwards. 
   \label{fig:masslossratio2}
}
\end{figure}

Figure~\ref{fig:central}  shows the 
     density, CR pressure,
and the ratio of CR pressure to the combined thermal and CR pressure
defined as $\epsilon=P_{CR}/(P_{CR}+P_{TH})$
    for
model ZC at $t=2.2$\,Myr in the disk where most
  mass is. The outflow is dominated by CR pressure with
  $\epsilon>0.5$, but both CR and thermal pressure keep pushing the
  disk gas to the side, pressing it to the midplane, instead of
  entraining it in the outflow. The compressed, high-density gas at
  the midplane is where most dispaced mass is, dominated by thermal
  pressure. 

We also ran a simulation ZI with a CR diffusion coefficient 
   $\kappa_{CR}=3\times10^{26}$\,cm$^{2}$\,s$^{-1}$ an
order of magnitude smaller than our fiducial value, 
to be compared to our model ZC. 
  \citet{SalemBryan2014} showed that mass-loading is more efficient
  with their smallest  CR diffusion coefficient,
  $\kappa_{CR}=3\times10^{27}$\,cm$^{2}$\,s$^{-1}$, compared to values
  one or two orders of magnitude higher that they also tested, 
because CRs remain in higher-density regions longer to exert pressure on the
  gas. (We chose 
  $\kappa_{CR}=3\times10^{27}$\,cm$^{2}$\,s$^{-1}$ as our fiducial
  value, based on their results.)


CRs indeed diffuse more slowly
in model ZI than model ZC, so they remain 
in the central few hundred parsec region of the disk longer, instead of
diffusing out into
the cavities created by the outflows. 
     Nevertheless,
the total mass
moving in model ZI
    remains similar to that in model ZC, with
 $3.5\times10^{6}$~M$_{\odot}$ at
$v_{z}>10$\,km\,s$^{-1}$ 
and $3.0\times10^{6}$~M$_{\odot}$ at
$v_{z}=55$\,km\,s$^{-1}$ at $t=1.2$~Myr.

Figure~\ref{fig:masslossratio2} shows that there is 
    less than a factor of two
difference in the total mass moving upward in outflows with or
without CRs, with varied diffusion coefficients. 
All the extra mass in CR-driven outflows is accelerated by the time
the main bubbles blow out at $t=0.2$--0.4\,Myr. The difference in loaded
mass between CR-driven and thermal-pressure driven outflows decreases as mass accelerated earlier by CRs moves off the computational grids. CRs no longer seem to
entrain mass in the outflows. 


We define the mass-loading ratio as the ratio of the amount of 
mass moving upward
to the star formation rate. The total mass in the
outflow is not likely to increase for $t>2.2$\,Myr (see
Figure~\ref{fig:central} and
Figure~\ref{fig:masslossratio2}), 
   so we calculate the mass-loading ratios using the amount of
   accelerated mass at $t=1.2$\,Myr before a large part of it leaves the computational grid. 
If we assume an instantaneous starburst of
$10^{9}$~M$_{\odot}$, the mass-loading 
     ratio is only 0.003--0.004 
while if we assume continuous star formation at a rate of
100--500~M$_{\odot}$\,yr$^{-1}$, we get mass-loading ratios of $\ll$0.006--0.03. 

We find that only slightly more
gas is entrained from the disk and accelerated with CR
pressure
    than with thermal pressure alone.
With a single star formation site at
the center of a disk, there is
    little substantive difference in 
SN driven outflow dynamics with or
without CR pressure.  
This result holds despite our assignment of 30\% of mechanical
energy to CRs, choice of the assumption of isotropy in CR diffusion, and
neglect of CR cooling.


%

In our simulations, the amount of mass that gains enough energy to
         reach 
the halo is determined at the time of blowout: this is
what we expect for feedback by a classic thermal pressure driven wind
based on previous numerical studies \citep[]{MF99, Fujita2009}.  Our
simulations suggest that in a large, ultraluminous starburst galaxy
with a single 
    major
SF site at the center, CRs do not seem to play any
significant role in removing gas from its dense disk
and massive halo potential.

\section{Missing Physics}
We neglected many pieces of physics for our study, 
    including
anisotropic diffusion and streaming of CRs, magnetic
fields, CR 
    spectrum
spectral evolution and cooling, and realistic 
star formation distributions.
We justify our choice to ignore the first five pieces of CR related physics in
the list above by noting that our aim was to compute the maximum effect
of CR pressure in our model 
galaxy 
with a single central SF site, 
within our limited computation time.  Recent simulations with
isotropic or anisotropic CR diffusion and CR streaming with or without
MHD have shown that the assumption of isotropic diffusion leads to the
highest mass-loading rate in a given numerical setting for a
galaxy \citep[e.g.][]{Wiener2016, Ruszkowski2016}.

On the other hand,
the soft, ultrarelativistic, 
polytropic equation of state
with $\gamma_{CR}=4/3$ that we assume is likely to reduce the effects of CR
pressure compared to the stiffer CR fluid expected from a larger spectral index
for the CR energy distribution \citep[]{Jublegas2008}. However, our
simulations show that 
the overall outflow dynamics 
     varies little
whether it is driven only by thermal
pressure or by both CR and thermal pressure, 
once the thermal pressure driven
superbubble overtakes the low-density CR driven bubble as shown in Figure~\ref{fig:big}.
In addition, our simulations show that the 
results 
     do not depend significantly on the strength of CR diffusion. 

We assumed a single SF site at the center of our disk to reproduce
our previous results \citep[]{Fujita2009} but clearly this is an
oversimplification.   We solved for the
evolution of a single, centrally concentrated starburst bubble based
on superbubble dynamics in a single OB
association: stellar winds create a hot, low-density cavity in the
interstellar medium, and repeated SNe excavate a larger hole as
they sweep the gas into a thin, dense shell \citep[]{tomisaka86,
  ML1988}. A single energy source transfers energy to the surrounding
rarefied interior gas
more efficiently than multiple smaller clusters or single SNe spread
over a large region in a disk \citep[]{Fragile2004},
because SN thermal energy dumped in high-density gas is vulnerable to
radiative cooling. This means in our simulations, we are computing the maximum effects of SN 
thermal pressure
   on accelerating gas.
However,
a single superbubble sweeps the ISM but pushes most of it to the sides,
as it blows out, so that only a small amount of swept-up gas is unbound from the
disk.  
Although CRs that quickly diffuse
isotropically through the disk and into the halo can further accelerate this small amount of
swept-up material, our simulations
  show that they do not entrain mass from elsewhere in the
  upper atmosphere, as seen in models presented by, for example,
  \citet{SalemBryan2014} and \citet{Girichidis2016}. 
As a result, there is virtually no difference between mass-loading rates of outflows
driven only by SN thermal pressure and of outflows driven both by SN
thermal pressure and CR pressure.

Because our initial conditions do not include the effects of star
formation feedback in forming a thick gas disk (the Lockman and
Reynolds layers in our own galaxy, for example) our model does not
have an extended moderate-density upper atmosphere to provide mass for
CR acceleration as occurred in the earlier models.  

The tunnel to
vacuum provided by our single, central superbubble may also play a
role in reducing the effect of CR acceleration on other parts of the disk.
However, recent
simulations of CR driven winds that find high mass-loading rates
treat star formation over the entire disk wherever the gas density exceeds a certain
threshold density \citep[]{Uhlige012,Hanasz2013, Booth2013,
  SalemBryan2014,Simpson2016, Girichidis2016, Martizzi2016,
  Pakmor2016,Ruszkowski2016}. CR pressure won't manage to unbind
rotationally supported, orderly gas in a disk, but we suspect it might 
have a bigger impact on gas
locally disturbed by superbubbles.
We plan to address the dependence of
mass-loading on star formation distribution in future work. 
Escaping CRs may also decelerate infalling gas in a
halo, or even accelerate it outward, thus inhibiting it from feeding
further star formation in a disk. 

\section{Conclusions}

   Our goal in this work was to answer the question of whether cosmic rays could
   enhance the ability of strong stellar feedback to drive galactic outflows
   in massive galaxies just below the mass where AGN feedback begins
   to dominate. To answer this question,
we presented the results of 3D, hydrodynamic
simulations of galactic outflows driven both by SN thermal pressure
and CR pressure with isotropic CR diffusion in an
ultraluminous galaxy
   with mass $5 \times 10^{12}$~M$_{\odot}$.
We modeled a single SF site at the center of the disk, and in this
setting, we computed the maximum effect of CR pressure on driving an
outflow by neglecting several pieces of physics. 
CRs diffuse quickly through the disk and into the halo. They can drive a
low-density bubble 
     that expands beyond the shell swept up by the
thermal pressure driven bubble. However, we find that the overall outflow
dynamics are quite similar whether the outflow is driven by thermal
pressure alone or by the combination of
thermal and CR pressure, yielding a negligible difference between the mass-loading
rates. Running simulations with varying CR
diffusion coefficients 
did not change this result.  

CRs appear to make little difference
in this luminosity range, 
   producing outflows with almost as little mass-loading as thermal
   driving alone.
       The path to vacuum opened up in the models by
       well-resolved thermal gas allows the CRs to escape easily,
       something that many previous models have not treated in detail.

Previous studies on SN feedback in
galaxies have shown that dark matter is key to controlling whether a
bubble blows out or blows away \citep[e. g.][]{MF99}, and likewise, dark matter may be key to the
development of CR-driven outflows. 
However,  we also speculate that simulations with 
multiple star forming regions distributed in space and time may
change the result because CRs can entrain and accelerate extended, turbulent, moderate to
low density upper atmosphere created by previous generations of star
formation more efficiently than rotationally supported disk gas bound
in the galactic potential. 

\section*{Acknowledgements}
M-MML was partly supported by NSF grant AST11-09395. We acknowledge
use of the Fujitsu FX10 at Shinshu University, the Cray
XC30 at the NAOJ, and resources from NSF XSEDE under grant TG-MCA99S024.   
Computations described in this work were performed using the
publicly-available \texttt{Enzo} code (http://enzo-project.org), which is
the product of a collaborative effort of many independent scientists from
numerous institutions around the world.  Their commitment to open science
has helped make this work possible.




\bibliographystyle{mnras}
\bibliography{akimi}{ULIRG2016} 







\bsp	
\label{lastpage}
\end{document}